\documentclass[preprint,12pt,authoryear]{elsarticle}

\usepackage{amssymb}

\usepackage{xcolor}
\usepackage{amsmath}

\usepackage{pdfpages}

\journal{TBD}

\begin{document}

\begin{frontmatter}



\title{A Clinical Instrument to Measure Patient Anecdotes in Clinical Trials} 


\author{} 

\affiliation{organization={Miller, Hyslop, and Decker},
            addressline={PO BOX 1233}, 
            city={Lyons},
            postcode={80540}, 
            state={Colorado},
            country={USA}}

\begin{abstract}
Clinical trials assessing neurological treatment are challenging due to the diversity of brain function, and the difficulty in quantifying it. Traditional treatment studies in epilepsy use seizure frequency as the primary outcome measure, which may overlooking meaningful improvements in patients' quality of life. This paper introduces the Clinical Instrument for Measuring Patient Anecdotes in Clinical Trials (Clinical IMPACT), a novel tool designed to capture qualitative non-seizure improvement across neurological domains.

The Clinical IMPACT incorporates open-ended inquiries that allow participants or caregivers to identify and select anecdotal evidence of their most significant treatment benefits. A blinded panel of experts ranks these anecdotes, facilitating a rigorous statistical analysis using the Wilcoxon Rank-Sum Test to detect treatment efficacy. The approach is resistant to type 1 error, yet comprehensive in its ability to capture real-world effects on quality of life.

The potential of the Clinical IMPACT tool to enhance sensitivity while also providing qualitative insights that can inform patients, healthcare providers, and regulatory bodies about treatment effects makes it important to consider in any neurological trial. We describe how it can be used in epilepsy, and advocate for its inclusion as a key secondary endpoint to provide a perspective on non-seizure outcomes, which have previously been challenging to measure, let alone to interpret, even when the clinical trial is positive.\end{abstract}


\end{frontmatter}



\section{Introduction}
\label{sec1}

Epilepsy remains a significant public health challenge, affecting millions worldwide \citet{who2019} and 1.1\% of self-reporting adults in the US \citet{kobau2023}. Conventional epilepsy treatment studies use seizure frequency as the primary outcome measure because it is quantifiable and clinically meaningful. Although seizure counts have traditionally been the standard accepted by regulatory bodies, they cannot identify other ways in which a therapeutic intervention may improve a patients' quality of life. Patient groups have argued for the importance of meaningful non-seizure endpoints, and advocated for their inclusion in trials \citet{hecker2024}.

The complex and varied functioning of the human brain presents a substantial obstacle in assessing non-seizure outcomes, especially across age groups. Conventional assessments have different strengths regarding the applicable age range, neurological domains tested, sensitivity to change, susceptibility to training on repeat testing, or administration efficiency. Existing measurement instruments typically lack adequate sensitivity for detecting significant changes in non-seizure-related outcomes across the entirety of the developmental spectrum. Consequently, investigators may observe improvements that suggest potential treatment efficacy, but which are not captured in structured tools. The most sensitive instrument to change is the Clinical Global Impression of Change (CGI, or CGI-C) scale, which allows participants to report "very much" or "minimal" improvement or worsening, but it provides no additional color regarding how or why the participant chose that term to describe their treatment response. Consequently, even a positive outcome based on the CGI can be difficult to understand clinically and pragmatically.

To address these limitations, we propose a novel method for data collection within randomized clinical trials. It has broad application, but we focus here on how it can be implemented for efficacy assessment in epilepsy treatment. This method incorporates a series of open-ended qualitative inquiries across key functional domains, including cognitive, communication, emotional and behavioral, social, motor, sleep, and overall quality of life (QOL). This tool prompts the participant to reflect on the most significant improvement within these domains, and then select the single anecdote that is most meaningful to them due to its effect on their overall quality of life. A blinded panel of experts (which can include epilepsy specialists and/or patient/family representatives), subsequently reviews and ranks these anecdotes to provide a hierarchical list of perceived quality-of-life improvements. This approach results in a blinded ordinal data set that can be analyzed using the Wilcoxon rank sum test.

This approach, which we descriptively call the  "Clinical Instrument for Measuring Patient Anecdotes in Clinical Trials" (or "Clinical IMPACT") can quantify treatment effects with enhanced sensitivity, as well as provides qualitative insights about the results, which are valuable to patients, treating physicians, regulatory bodies, and payors who wish to understand how treatment affects overall quality of life. By illustrating how study subjects have responded to an intervention, Clinical IMPACT provides a comprehensive evaluation framework that allows characterization of non-seizure endpoints in a statistically powerful way. Since it is applicable to all ages and developmental levels, we advocate its inclusion in all future epilepsy clinical trials as a first key non-primary endpoint.

\section{Methodology}
\label{sec2}

\subsection{Overview of the Data Collection Tool}
\label{subsec2-1}

The methodology behind Clinical IMPACT creates a qualitative self-reporting instrument that measures non-seizure improvement in an epilepsy treatment trial, which patient advocates say is important. It is valid for any age range, suitable for individuals with varying neurological morbidity, is not susceptible to treatment effect, and focuses on biggest successes rather than asking if participants have achieved (or failed) to reach a predefined goal. As a result, it is capable of providing meaningful insights in any population with epilepsy.


\subsection{Administration of the Instrument}
\label{subsec2-2}

The Clinical IMPACT tools consists of a series of structured questions that encourage participants (or the reporting caregiver) to reflect on significant improvements observed in any domain since beginning treatment. The questions - which explicitly elicit changes in cognition, communication, emotional and behavioral regulation, social, motor, and sleep functions - are chosen to ensure exhaustive consideration of neurological function, with answers checked to allow meaningful comparison between study participants that an expert panel can understand and rank.

The purpose of covering individual domains is to help participants be comprehensive in their recall. Once every domain is considered and discussed, the participant is asked to select the \textit{single biggest improvement} they have experienced since the initiation of treatment. 

The administration of this data collection tool must be conducted by any trained member of the study team, including Principal Investigator (PI), neuropsychologist, nurse, or study coordinator. While these team members will be responsible for reviewing and facilitating the process, the questions result in an open-ended, self-reported item that allows respondents to characterize their experience during the trial. As a result, the individual administering the instrument should be enthusiastic about any improvement the reporter identifies, but should provide the quality check described in the next section.

Participant recall for this instrument is expected to benefit from other instruments (such as the Vineland Adaptive Behavior Scales) being completed beforehand. Therefore, the Clinical IMPACT should be assessed as the final outcome measure (chronologically on the study day; not necessarily in the statistical analysis plan), when possible. This does not affect the validity of the conclusions, but does increase its sensitivity.

For studies that include an assessment of the Clinical Global Impression of Change, the Clinical IMPACT should be elicited \textit{after} the CGI, since it can provide color or context to how the participant answered the CGI question and should be congruent with it (meaning that the anecdotes that are shared should support the direction and magnitude of change they report on the CGI).

Timing during the period of treatment is also important to consider. Since the therapeutic effect of most treatments is expected to be most clear after steady state is reached and time has passed, the Clinical IMPACT should be administered on the \textit{last day of blinded treatment} (at the time of study roll-over to active treatment, for example). This allows the data set to reflect any benefits of treatment that may occur later in the treatment window. This incurs a small cost of lower recall for early improvements that are not sustained, but this is likely justified by the opportunity to see the aggregate effects of longer treatment. Nevertheless, because the instrument is not susceptible to training effects or other errors introduced by repeated measurement, it \textit{is} acceptable to perform the Clinical IMPACT assessment more than once during a treatment protocol. If there is only a single point of observation, however, then the most important time to assess Clinical IMPACT is at the conclusion of blinded treatment.

\subsection{Ensuring Quality of Responses}
\label{subsec2-3}

Once the participant identifies the most meaningful improvement, the test administrator must review the anecdote chosen by the participant to confirm that:
\begin{itemize}
    \item the item is anecdotal rather than general.
    \item the item includes an internal point of comparison for that participant (how it compares to their expectations, or what was "typical" prior to treatment).
    \item the item contains no proper names or other personally-identifiable information (PII).
\end{itemize}
 
These features are necessary for the blinded expert panel to compare them in a meaningful way, and to protect patient privacy. Since an open-ended text field in the study database is required for the anecdote to be captured, direct elimination of PII at the time of the study visit is critically important.

Anecdotal means that the response describes \textit{one specific point in time}. The point in time need not be included (or even known) as long as it is specific. In circumstances where parents provide a general observation, the trained administrator should reply with enthusiastic support of any positive report, and follow up with a solicitation for the one point in time they got the strongest sense of that improvement, or which best demonstrates that idea. 

For instance, rather than stating "I think she's been more aware of her surroundings", a more informative response would be, "One morning, I noticed that he smiled when his favorite song came on." Specific examples allow for a clearer understanding of the impact treatment has had on the participant's life, which are necessary for subsequent inter-subject comparison by the expert panel.

An example dialogue between a PI performing the assessment and a study subject is shown in Figure \ref{fig1}.

\begin{figure}[t]
\begin{itemize}
    \item \textbf{PI}: (\textit{reading the verbiage directly from \ref{app1}}) Out of everything that has happened since starting treatment, please choose one story that best demonstrates how the treatment affected the participant's life (aside from seizures).

    \item \textbf{Subject}: I can think more clearly!
    \item \textbf{PI}: (\textit{recognizing that this is a general observation and not anecdotal}) "We are happy to hear about that kind of improvement!
    \item \textbf{PI}: (\textit{(providing support and endorsement for the improvement, but focusing on the single strongest supportive example)}) Can you tell me about one specific time that you noticed that most clearly? Or give a specific example that best demonstrates what you mean?
    \item \textbf{Subject}: "Oh; sure. That's easy. When I went to the store last week I was able to recall all four items I needed.
    \item \textbf{PI}: (\textit{seeking an internal baseline or point of comparison}) That is such welcome news. How was that different than usual? What would normally happen?
    \item \textbf{Participant}: Normally I need to check my list even if it's just one item
    \item \textbf{PI}: Perfect. It is easy to understand how meaningful that would be in your life
\end{itemize}

\caption{Example of PI-participant interaction when a general observation is provided instead of an anecdote, and making sure a point of comparison exists}\label{fig1}
\end{figure}

\subsection{Blinded Expert Panel Review and Ranking}
\label{subsec2-4}

Once anecdotes are collected a blinded panel of experts (which may include clinical epilepsy specialists and/or patient/family epilepsy advocates), reviews the anecdotal reports provided by participants. The blinding is a necessary feature to guarantee that the chosen ranking is scientifically valid, and therefore statistically quantifiable and interpretable. As a result, the clinical experts on the panel should not themselves be PIs for any site.

The expert panel reviews the single anecdote chosen by each participant as their single biggest improvement, then ranks all anecdotes (one per participant) against the others. This ranking process results in a single list with a value from "n" (the  study population size; the anecdote that demonstrates the most clinically meaningful improvement) to 1 (which is the least clinically meaningful improvement). The analysis is statistically valid whether the ranking is forced (no two anecdotes are allowed to have the same rank) or whether ties are allowed (this circumstance is handled by giving all "tied" anecdotes the same average rank). The validity of the instrument does not depend on \textit{when} the ranks are assigned. As a result they can be ranked when each subject reaches the end of blinded treatment, or in a batch at the end of the study. This feature of the Clinical IMPACT means that interim analyses are possible (and scientifically valid).

The ranking process allows for a comprehensive assessment of the perceived significance of each anecdote. Recognizing the subjective nature of this evaluation, committee members will be instructed to consider the context and emotional weight of each anecdote (using only the text of the anecdote) while assigning rankings.

This will require the expert panel to choose an order for anecdotes in very different domains with impacts that are therefore hard to compare, but the statistical conclusions will be insensitive to what is decided when the order is close.

The suggested governance of the expert panel is to have one clinical expert serve as chairperson who elicits thoughts, input, and perspective regarding each anecdote from the other panel members, but who is then responsible for the final rank assignment. Other governance models can also be chosen without reducing the statistical confidence of a positive trial, but because the assigned ranks are susceptible to noise, a dysfunctional governance model could result in a false negative conclusion as well as other pragmatic inefficiencies.

\subsection{Statistical Analysis}
\label{subsec2-5}
\subsubsection{Methodology}
Each participant is ranked according to their perceived level of clinically meaningful improvement as summarized by their associated anecdote, and the statistical analysis aims to test for a significant difference in the level of improvement between the two treatment groups based on this ranking. The Wilcoxon Rank-Sum Test\cite{brunner2018rank} is well suited to demonstrate an effect in this setting, as  
the test statistic and an intuitive directional treatment effect, the relative effect, can be computed from ranks alone. This is crucial because the Data Collection Tool produces data that are themselves ranks;  there are no explicit underlying numeric or ordinal measurements available from which the ranks are derived. We comment further on this nuance after briefly explaining the intuition behind the Wilcoxon Rank-Sum Test and our approach to statistical analysis and its implementation. 

For two treatment groups, $A$ and $B$, with $n_A$ and $n_B$ participants, the ranks that were assigned by the expert panel to all participants combined are summed over the participants within each group, resulting in the rank sums $R_A$ and $R_B$. The Mann-Whitney statistics $U_A$ and $U_B$, which count the number of times (in pair-wise match-ups) that participants in group $A$ are ranked higher than participants in group $B$, and vise versa in the case of $U_B$, depend on the data only through $R_A$ and $R_B$ respectively. The test statistic is derived from $U_A$ and $U_B$ which can be computed with ranks alone. 

Intuitively, if $U_A$ is large,  so that participants in  group $A$ out rank participants in group $B$ more often than they rank lower, there may be significant evidence that the anecdotes generated by these participants indicate more meaningful improvement. Significance is determined by  testing the null-hypothesis
\[
H_0:=\textrm{\textbf{ The distribution of the ranks is the same across the groups.}}
\]
against the alternative
\[
H_A:=\textrm{\textbf{ One group tends to have higher ranks than the other.}}
\]
and rejecting $H_0$ in favour of $H_A$ when the Mann-Whitney $U$ statistic,  $U= \min\{U_A, U_B\}$ is small.  

For small values of $n_A$ and $n_B$ (ranging between 1 and 25), and if there are no tied ranks, exact $p$-values are known and have been tabulated for every possible value of $U$.   The Wilcoxon Rank-Sum Test is non-parametric and can in principle be used accurately whenever it can be calculated from the data (no minimum sample size). For sample sizes outside of the tabulated range, or if there are tied ranks,  normal approximation is frequently used to compute $p$-values \cite{brunner2018rank}.

If $H_0$ is rejected in favor of $H_A$ there is evidence that one groups tends to have higher ranks than the other, and it is useful to report a measure of treatment effect that clarifies the direction of the effect. 

A common treatment effect, the relative effect, $\hat p_A$ (or $\hat p_B$) \cite{brunner2018rank} is a consistent estimator of the probability that a randomly chosen anecdote generated by group $A$ out ranks a randomly chosen anecdote generated by group $B$ (or vice versa). The treatment effect $\hat p_A$ (and also $\hat p_B$) can be estimated using only ranks and are in fact re-scalings of $U_A$ (respectively $U_B$).  We suggest reporting both the $p$-value of the Wilcoxon rank sum test, and the larger of $\hat p_A$ and $\hat p_B$ as a measure of effect size.

\subsection{Implementation}
Most implementations of the Wilcoxon Rank-Sum Test (for example, the R function wilcoxon.test) take as their input ordinal data, and compute the ranks required to implement the test based on this data. 

To carry out the statistical analysis, we wrote an R function that implements the Wilcoxon Rank-Sum-Test starting with the ranks. The code is available upon request.

An example, suppose treatment group $A$ and treatment group $B$ consist of $5$ and $6$ participants respectively so that $n_A=5$ and $n_B=6$. The ranks of the participants in treatment group $A$ are  $\{11, 6, 8, 5, 9, 10\}$ and the ranks in of the participants in treatment group $B$ are $\{2, 4, 3, 1, 7\}$. Then $R_A = 49$, $R_B=17$, and the null-hypothesis is rejected with a $p$-value of $0.018$, based on the normal approximation methodology implemented in our R function and taken from \cite{brunner2018rank}. An estimate for the treatment effect, that a randomly chosen element from group $A$ is ranked higher than a randomly chosen element from group $B$ is $\hat p_A = 0.93$. Since the $p$-value of the test is less than $.05$ there is significant evidence that the anecdotes generated by the two treatment groups do not have the same distribution of ranks. Since the test shows significance and $\hat p_A = 0.93>0.5$ there is significant evidence that anecdotes generated from group $A$ tend to be more highly ranked. The assumptions of the Wilcoxon Rank-Sum Test are satisfied under realistic behavior of the expert review panel, namely that ranks are consistently assigned so that the relative rank (which one is higher and which one is lower) assigned to two anecdotes does not depend on which anecdotes are ranked along side them.

\section{Discussion}

\label{sec3}

\subsection{Overview of the Method}
\label{subsec3-1}

We have described the Clinical IMPACT tool; an instrument for collecting qualitative and anecdotal data that can provide statistical confidence that a treatment effect is present. We have tailored the method for use in clinical trials of epilepsy and describe how sensitivity of the outcome can exist, even if the specific domain of functional improvement is unknown or difficult to anticipate at the time of trial design. By focusing on anecdotes that are personally meaningful to participants (or their reporting care providers), we can capture nuanced experiences that traditional quantitative measures may overlook. This qualitative approach enriches the data collected, allowing a more holistic understanding of treatment effectiveness.

\subsection{Statistical Characteristics of the Method}
\label{subsec3-2}

Using the Clinical IMPACT as described has an extremely low risk of positive bias due to the blinding of the expert panel. As a result, the chance of a false positive study should be extremely close to a theoretically perfect design where the risk of a false positive equals the selected \textit{p-value}. As a result, regulatory agencies are expected to embrace the approach: they only review data that the sponsors interpret as positive, so the risk of type 2 (false-negative) data is not relevant to approval decisions.

A false negative result is possible, and the chance of seeing such a result is more difficult to estimate. It depends on recall by participants, difficulty in comparing and ranking anecdotes in qualitatively different domains (see \ref{subsec3-4}), and having an interpretable anecdote submitted by each subject (see \ref{subsec2-3}). As a result, this outcome measurement is not straightforward to use as a model for purposes of power estimation during study planning.

\subsection{Advantages over Conventional Tools}
\label{subsec3-3}

The Clinical IMPACT tool does not require the instrument to anticipate the way in which function or quality of life might improve. This is an advantage over most neuropsychological assessments, which can assess different domains but may not elicit all aspects of function within that domain.

This evaluation tool also provides qualitative context to the quantitative statistical conclusion: the actual list with the assigned rankings is useful for patients to understand the real-world implications of treatment. Similarly: the anecdotes submitted help prescribing physicians, study sponsors, and health care payors understand the treatment effect beyond the primary endpoint \textit{p-value} generated by the study.

Finally, we note that even if the Clinical IMPACT fails to demonstrate statistical significance, the anecdotal data points may provide color regarding participant experiences during the treatment. The anecdotes may also provide guidance for subsequent trial design using the same intervention, thereby improving sensitivity of later trials.

\subsection{Criticisms of arbitrary or subjective ranks}
\label{subsec3-4}

One possible criticism of the Clinical IMPACT tool is the subjective (and possibly arbitrary) nature of the ranks assigned by the expert panel to anecdotes that are in widely disparate functional domains.

This concern is valid: the subjective nature of ranking anecdotes is influenced by how the expert panel estimates utility - an intrinsically personal process.  However, the most important consideration is that the truth of the subjectivity of the ranks \textit{cannot indict the statistical certainty of a positive result}. This is because arbitrary assignment can create noise, and the only effect of noise in a well-designed and executed study is to increase the chance of a false \textit{negative}. There is no mechanism by which it can lead to a false \textit{positive}, and therefore any positive result remains valid.

It is also critical to recognize, however, that any reviewer or critic can re-order and re-analyze the results themselves. This process would be unblinded, and therefore not a statistically measurable or actionable exercise, but it would be trivial to perform a kind of sensitivity analysis to understand how different ranks would have affected the result (such as in a critical review).

Although this exercise is possible, it is unlikely to affect the results of a highly effective treatment. This is because the test statistic derived by the Clinical IMPACT is influenced by the appearance of actively treated patients near the top and placebo-treated patients near the bottom. How patients within those cohorts are ranked relative to each other will influence the test statistic in a smaller way.

By articulating these limitations, we aim to foster appropriate use of the Clinical IMPACT tool, as well as encourage discussion, suggestions, or future studies regarding how this process can be improved or extended.

\section{Conclusions}
\label{sec4}

This study introduces a novel qualitative self-reporting instrument called the Clinical IMPACT tool. We discuss how to customize it to assess the multifaceted impacts of epilepsy treatment across various functional domains. Unlike existing neuropsychological instruments, which often lack sensitivity to subtle qualitative changes, this instrument is fully sensitive to such improvements through its focus on detailed anecdotes provided by caregivers and family members.

By incorporating a series of open-ended inquiries addressing the domains of cognition, communication, emotions and behavior, social, motor, sleep, and overall quality of life, the instrument captures the most meaningful effect of treatment  on participants' quality of life.

The innovative methodology includes a blinded panel of experts who rank the reported anecdotes, enabling rigorous statistical estimation of the probability that a treatment effect is present. This dual capability of sensitivity and statistical rigor allows a comprehensive understanding of treatment efficacy, moving beyond conventional measures that ofter overlook significant patient experiences.

We advocate for the integration of the Clinical IMPACT in future epilepsy clinical trials, as it provides both qualitative depth and quantitative reliability. This approach aims to enhance therapeutic strategies and improve the quality of life for individuals living with epilepsy, ultimately bridging the gap between subjective experiences and objective assessments in clinical practice.

\appendix
\section{Case Report Form}
\label{app1}

\includepdf[pages=1-5]{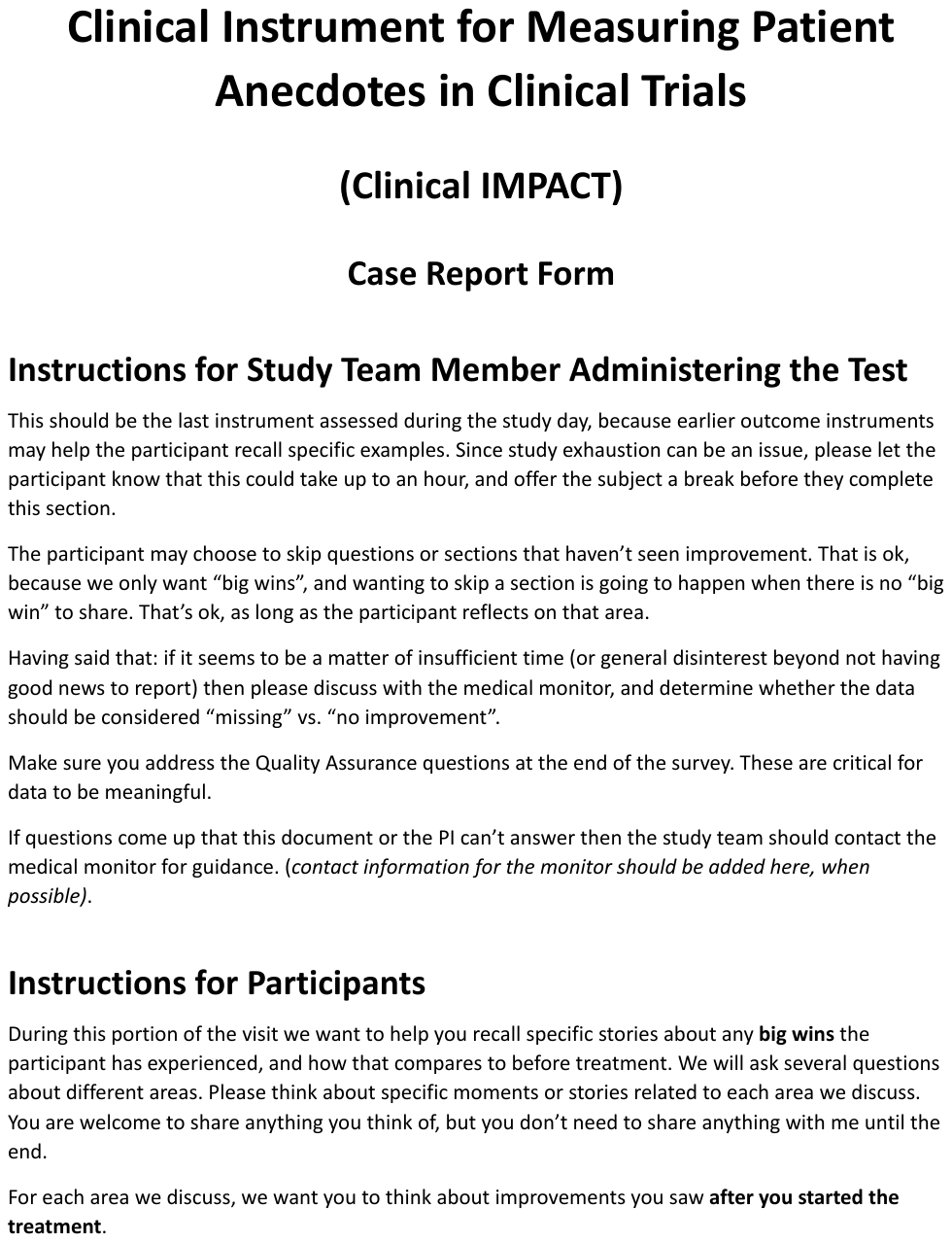}




\begin{thebibliography}{00}


\bibitem[(Brunner et al. 2018)]
{brunner2018rank}

    \bibitem[(Hecker 2024))]{hecker2024}
          Hecker J, Conecker G, Chapman C, Hommer R, Ludwig NN, Sevinc G, Te S, Wojnaroski M, Downs J, Berg AT. ,
          \textit{Patient-advocate-led global coalition adapting fit-for-purpose outcomes measures to assure meaningful inclusion of DEEs in clinical trials.},
          Ther Adv Rare Dis. 2024 Jun 22;18:26330040241249762. doi: 10.1177/26330040241249762. Erratum in: Ther Adv Rare Dis. 2024 Oct 4;5:26330040241277869. doi: 10.1177/26330040241277869. PMID: 38911512; PMCID: PMC11193340.,
          2024.
        

    \bibitem[(Kobau, 2023)]{kobau2023}
          Kobau, R., Luncheon, C., \& Greenlund, K. ,
          \textit{Active epilepsy prevalence among U.S. adults is 1.1\% and differs by educational level},
          National Health Interview Survey, United States, 2021. Epilepsy \& behavior : E\&B, 142, 109180. https://doi.org/10.1016/j.yebeh.2023.109180,
          2023.
    
    \bibitem[(WHO, 2019)]{who2019}
        WHO ,
        \textit{Epilepsy: a public health imperative},
        Geneva: World Health Organization,
        2019.        
    



\end{thebibliography}



\end{document}